\documentclass[epsfig,12pt]{article}
\usepackage{epsfig}
\usepackage{graphicx}
\usepackage{relsize}
\usepackage{amsmath,mathtools}
\usepackage{amsfonts}

\newcommand{\beq}{\begin{equation}}   
\newcommand{\eeq}{\end{equation}}
\newcommand{\beqn}{\begin{eqnarray}}   
\newcommand{\eeqn}{\end{eqnarray}}

\newcommand{\gsim}{\lower.7ex\hbox{$
\;\stackrel{\textstyle>}{\sim}\;$}}
\newcommand{\lsim}{\lower.7ex\hbox{$
\;\stackrel{\textstyle<}{\sim}\;$}}

\begin{document}

\begin{titlepage}
\begin{flushright}
FTPI-MINN-15/37, UMN-TH-3446/15\\
July 30/2015
\end{flushright}

\vspace{16mm}

\begin{center}
{  \bf{\large  Impact of Axions on Confinement in Three and Two \\[2mm]Dimensions}}
\end{center}

\vspace{6mm}

\begin{center}

 {\large 
M. M. Anber$^a$ and M. Shifman$^{a,b}$}
\end {center}

\begin{center}

$^{a}${\em Institut de Th\'eorie des Ph\'enom\`enes Physiques, EPFL, CH-1015 Lausanne, Switzerland.}

 \vspace{2mm}
 
$^b${\em William I. Fine Theoretical Physics Institute, University of Minnesota,
Minneapolis, MN 55455, USA}\,\footnote{Permanent address.}

\end {center}


\vspace{10mm}

\begin{center}
{\large\bf Abstract}
\end{center}

\hspace{0.3cm}

In this paper we discuss well-known three- and two-dimensional
models with confinement, namely, the Polyakov compact electrodynamics in 3D and two-dimensional $CP(N-1)$ sigma model,
and reveal   changes in the confining regimes of these model upon adding the axion field.

In both cases the addition of axion has a drastic impact. In the $CP(N-1)$ model the axion-induced deconfinement was known previously, but we discuss a new feature not considered in the previous publication.
	
\vspace{2cm}

\end{titlepage}

\newpage

\section{Introduction}

In this paper, we discuss well-known three- and two-dimensional
models with linear confinement, namely the Polyakov compact electrodynamics in 3D and two-dimensional $CP(N-1)$ 
sigma model, 
and reveal changes the confining regimes undergo if one adds the axion field.
Historically the first was the Polyakov model \cite{Polyakov:1976fu}. A confining string
in this model appears due to the fact that a mass is generated for the dual photon 
by the instanton-monopole contribution. As a result, domain walls (more exactly, domain lines) appear.
They play the role of the  confining strings. 
 The dual photon-axion mixing
drastically changes the domain line composition resulting in a``leackage" of a part of the electric flux 
of the probe charges into the Coulomb regime. We also show that the same phenomenon takes place 
in pure Yang-Mills theory on $R^3\times S^1$ with small circumference of $S^1$, provided an
appropriate deformation is added.  

As was  discussed in \cite{Gorsky06}, in 2D (nonsupersymmetric) $CP(N-1)$ model 
axions, being added in a conventional way, result in deconfinement.
Here we show that  formerly stable mesons start decaying into ``quark-antiquark" pairs
with exponentially suppressed probabilities at large $N$,
$$ 
w\sim \exp\left( -cf^{3/2}_aN^{1/4}\right),
 $$
where $f_a$ is the axion constant (dimensionless in two dimensions), and $c$ is a numerical coefficient.  Thus, at $N=\infty$
the theory is still confining, but at $1\ll N<\infty$ a ``weak" deconfinement occurs. 
At $N\sim 2$ one can expect full-blown deconfinement,
much in the same way as it occurs in supersymmetric model \cite{Witten78}. 

\section{Polyakov's Confinement in 2+1 Dimensions}\label{ChapterPolyakovConfinement}

We start from a brief review of Polyakov's compact electrodynamics in three dimensions,
and outline confinement mechanism in this model. Then, we show that inclusion of axion
completely destroys linear confinement of electric charges.
Polyakov's model of color confinement \cite{Polyakov:1976fu}
was historically the first gauge model
in which linear confinement of probe electric charges was analytically 
established in $2+1$ dimensions.

\subsection{Preliminaries}

To make QED compact,  
Polyakov suggested to embed it in the  Georgi--Glashow
model  
in 1+2 dimensions \cite{Polyakov:1976fu}. Conventional 't Hooft-Polyakov mono\-poles \cite{'tHooft:1974qc,Polyakov:1974ek} have to be reinterpreted as instantons
in the Euclidean version of the model (we will refer to them as to monopole-instantons).
To begin with, we will briefly outline the Polyakov mechanism limiting ourselves to the
$SU(2)$ case.

The Lagrangian of the Georgi--Glashow model \cite{Georgi:1972cj} in 2+1 dimensions includes gauge fields
and a real scalar field, both in the adjoint representation of 
$SU(2)$. 
The Lagrangian of the model is obtained from Yang-Mills in four dimensions by reducing to 3D (see Section \ref{4D YM}),
\beq
{\mathcal L}= \frac{1}{4g_{\rm 3D}^2}\, G_{\mu\nu}^a\, G_{\mu\nu}^a
+\frac{1}{2} (\nabla_\mu\chi^a )(\nabla_\mu\chi^a )
- \lambda (\chi^a \chi^a - v^2)^2\,,
\label{14.1}
\eeq
where $g_{\rm 3D}$ is the 3D coupling constant and $\mu\,, \nu = 1,2,3$; 
the covariant derivative in the adjoint acts as
\beq
\nabla_\mu\chi^a =\partial_\mu \chi^a + \varepsilon^{abc} A_\mu^b\chi^c\,,
\label{14.2}
\eeq
and the Euclidean metric is $g_{\mu\nu} = {\rm diag}\, \{+1,+1,+1\}$.
It is understood that $\lambda \to 0$, thus the last term is just a shorthand for the
boundary condition of the $\chi$ field,
\beq
\left( \chi^a \chi^a\right)_{\rm vac} = v^2\,,
\label{14.3}
\eeq
where $v$ is a real positive parameter. One can always choose the gauge
in such a way that
\beq
\chi^{1,2} =0\,,\quad \chi^3 = v\,.
\eeq
Then, the third component of $A_\mu$ (i.e. $A_\mu^3$)
remains massless. At distances larger than $1/m_W$ the field $A_\mu^3$ acts as a {\em bona fide}  photon.
At the same time, the $A_\mu^\pm = \frac{1}{\sqrt{2}g_{\rm 3D}}\, \left(A_\mu^1\mp A_\mu^2
\right)$ components become $W$-bosons; they acquire a mass $m_W=g_{\rm 3D}v$. This is why the model is referred to as compact electrodynamics.

The classical equations of motion which follow from Eq.~(\ref{14.1})
(second order differential equations)
can be replaced by first-order  equations,
\beq
-\frac{1}{2g_{\rm 3D}}\,  \varepsilon_{\mu\nu\rho} G^a_{\nu\rho}
= \pm \nabla_\mu\,\chi^a\,.
\label{14.4}
\eeq
The monopole-instanton action is
\beq
S_{\rm inst} = 4\pi \, \frac{v}{g_{\rm 3D}}  \equiv4\pi \, \frac{m_W}{g_{\rm 3D}^2}\,.
\label{14.5}
\eeq

The monopole-instanton in the model at hand has four collective coordinates: three translational and one 
phase corresponding to the unbroken $U(1)$ subgroup of $SU(2)$. After integrating over the $U(1)$ collective coordinate, we obtain the instanton measure in the form
\beq
d\mu_{\rm inst} = {\rm const}\times m_W^3\,   d^3 x_0\, 
\exp\left(-S_{\rm inst}\right)\,.
\label{14.7}
\eeq
The validity of the quasiclassical approximation demands that
$v\gg g_{\rm 3D}$ and hence, $S_{\rm inst} \gg 1$. As a result, the 
instanton measure carries an exponential suppression.

\subsection{Compact electrodynamics}
\label{22}

The mechanism we are interested in is applicable at
distances $\gg m_W^{-1}$. Then the presence of the
$W$-bosons in the spectrum of the model is irrelevant,
and one can focus on ``massless" fields (the meaning of the quotation marks will become
clear shortly). There are two such fields: the photon and oscillation quanta  of $\chi^3$,
\beq
\chi^3 = v +\beta\,.
\eeq
In what follows, we will omit the isospace index 3 to ease the notation.
We will endow the $\beta$ field with a mass $m_\beta$ such that $m_W \gg m_\beta \gg m_\varphi$
(i.e. $\lambda\neq 0$, albeit small), see Eq. (\ref{14.33}).
Then it
plays no role and can be ignored in  what follows.
In three dimensions, the photon field
has only one physical (transverse) polarization.
This means that the photon field must have a dual description
in terms of one scalar field $\varphi$ of the angular type \cite{Polyakov:1976fu}.

In the absence of source (probe) charges, one can always use the so-called first order formalism.
Consider $F_{\mu\nu}$ to be an independent variable and implement
\beq
F_{\mu\nu}=\partial_\mu A_\nu - \partial_\nu A_\mu
\eeq
 via introducing a field $\varphi$ with the action
\beq
\Delta {\cal S}_\varphi = \int d^3 x \left( \partial_\mu\varphi \right) \epsilon_{\mu\alpha\beta} F_{\alpha\beta}\,.
\eeq 
Now, varying with respect to the photon field we arrive at
\beq
F_{\mu\nu} - \epsilon_{\mu\nu\alpha} \partial_\alpha \varphi - \epsilon_{\mu\nu\alpha} a\, \partial_\alpha \beta =0\,;
\label{11}
\eeq
see Eq. (\ref{20}) which explains the occurrence of the last term 
in the right-hand side, cf. (\ref{duality}).
As was mentioned, the $\beta$ field is sufficiently heavy and can be ignored in the
low-energy limit of compact electrodynamics. Therefore, one can ignore the last term in (\ref{11}) (much in the same way as in (\ref{duality})).
Introducing a static electric charge as a source term $J$, we conclude that
\beqn
F_{\mu\nu} &\propto& \epsilon_{\mu\nu\alpha} \partial_\alpha \varphi \,,\nonumber\\[2mm]
J_\mu &=& \partial_\nu F_{\mu\nu} = \partial_\nu  \epsilon_{\mu\nu\alpha} \partial_\alpha \varphi \,.
\eeqn
The latter equality is only possible because of singularities (vortices) in the angular field $\varphi$. 
For the standard minimal vortex,
\beq
F_{\mu\nu}
 = \frac{g_{\rm 3D}^2}{2\pi } \, Q \, \varepsilon_{\mu\nu\rho}\left( \partial_\rho\,\varphi\right) \,.
 \label{14.16}
\eeq
 The normalization of the $\varphi$ field is chosen in such a way that the values
$\varphi = 0, \pm 2\pi , \pm 4\pi, ...$ are identified, i.e., $\varphi$ is defined on $S^1$, the circle with unit radius.
Then, given the coefficient in (\ref{14.16}),  the minimal probe electric charge ($Q=1/2$ in the model at hand)
creates a minimal single winding vortex  of the $\varphi$ field.
The original energy functional reduces to 
\beq
{\cal E} =\frac{1}{2g_{\rm 3D}^2}\int \, d^2 x \left(\vec E^2 +B^2\right)=
\frac{g_{\rm 3D}^2}{32\pi^2}\int \, d^2 x \left[\left(\vec\nabla\varphi\right)^2 +
\dot\varphi^2\right]\,.
 \label{14.17}
\eeq
At this level, the dual photon field $\varphi$ is massless.
Instanton-induced interaction generates a potential for the $\varphi$ field, however,
\beq
{\cal L}_{\rm inst} =\frac{1}{2}\, \mu^3\, \exp (\pm i\,  \varphi )\,, \qquad \mu^3 \sim m^3_W \exp\left(-S_{\rm inst}\right)\,,
\label{14.23}
\eeq
where $\pm$ refers to monopole-instanton (anti-instanton).
Assembling the monopole-instanton and anti-instanton contributions,
we arrive at the following effective Lagrangian for the field
$\varphi$:
\beqn
{\cal L}_{\rm dual}
&=& \frac{\kappa^2}{2}\,\left(\partial_\mu\varphi\right) \left(\partial_\mu\varphi\right) + 
\mu^3\, \cos  \varphi  \,,\nonumber\\[2mm]
 \kappa &=&\frac{g_{\rm 3D}}{4\pi}\,.
\label{14.32}
\eeqn
This is the Lagrangian of the {\em sine-Gordon} model. 
The dual photon mass is readily calculable from (\ref{14.32}),
\beq
m_\varphi = \mu^{3/2}\,\kappa^{-1}\,,
\label{14.33}
\eeq
which is exponentially small. The potential in (\ref{14.32}) is $2\pi$ periodic, as expected.

\subsection{Domain line as a confining string}
\label{sec23}

The $2\pi$ periodicity of ${\cal L}_{\rm dual}$ and the mass generation in (\ref{14.33})
results in the existence of domain lines of the type \cite{ShBook}
\beq
\varphi = 2\left[{\rm arcsin}\,{\rm tanh}\left(  m_\varphi y
\right)+\frac{\pi}{2}
\right]\,,
\label{14.35}
\eeq
interpolating between $\varphi_{\rm vac} = 0$
at $y=-\infty$ and $\varphi_{\rm vac} = 2\pi$
at $y=\infty$, where $y$ is one of two coordinates 
in the two-dimensional $\{x,\,y\}$ plane. The transverse size of the 
domain line is obviously $\sim m_\varphi^{-1}$,
while its tension is 
\beq
T = 8\mu^{3/2}\,\kappa = 8 \, m_{\varphi}\,\kappa^2 \,.
\label{14.36}
\eeq
Note that this tension is much larger than $ m_\varphi^{2}$.

The above domain line is in fact a string that ensures
 linear confinement of the probe electric charges in compact electrodynamics. 
 Indeed, the necessary conditions for the topological defect
to be a string are: (i)  the defect is a one-dimensional object;
(ii) while traveling away from the defect in the transverse direction, 
at large distances, we should find ourselves in one and the same vacuum no matter in 
which direction we go. The first requirement is obviously satisfied for a long
domain line. The second requirement is also satisfied since
for the  compact field $\varphi$   we  have
physically the same vacuum on both sides of the  
domain line. 

For the linear regime to set up, the distance between the probe charges $L$ must be $L\gg m_\varphi^{-1}$.
The tension of this string is given in (\ref{14.36}).

\subsection{Axion's impact}
\label{Axion's impact}

If we introduce in (\ref{14.1}) an appropriately normalized vacuum angle $\theta$,
\beq
\Delta {S }_\theta = \frac{\theta}{16\pi^2} \, \epsilon_{\mu\nu\alpha} \, G_{\mu\nu}^a \left({\nabla_\alpha} \chi^a\right)\,,
\label{20}
\eeq  
then the instanton Lagrangian (\ref{14.23}) takes the form 
\beq
{\cal L}_{\rm inst} =\frac{1}{2}\, \mu^3\, \exp \left[ \pm i\, \left( \varphi + \theta\right)\right].
\eeq
It is obvious that the $\theta$ angle can be absorbed in $\varphi$ and completely disappears from the
physics in compact electrodynamics. This observation was first made by Polyakov in the 1970s.

This statement does not extend to the axion field, however, because adding the axion field ($\theta\to \theta + a$) introduces an extra dynamical degree of freedom. The Lagrangian (\ref{14.32}) now takes the form\,\footnote{It is assumed here that for given $\theta$ there are no other vacua entangled in the $\theta$
evolution. At strong coupling this need not be the case. For instance, in QCD with $N_f$ light flavors the $\theta$ dependence appears as $f(\theta/N_f)$, where the function $f$ is $2\pi$ periodic. This is due to the fact that
$N_f$ vacuum states are entangled  in the $\theta$
evolution. At $\theta = \pi, 3\pi, 5\pi$ one jumps from one vacuum to another. The assumption of a single vacuum involved  in the $\theta$
evolution is not important for our statement.}
\beq
{\cal L}_{\rm eff}
= \frac{\kappa^2}{2}\,\left(\partial_\mu\varphi\right) \left(\partial_\mu\varphi\right) + 
 \frac{f_a^2}{2}\,\left(\partial_\mu a \right) \left(\partial_\mu a \right) +
\mu^3\, \cos  (\varphi + a )\,,
\label{ququ}
\eeq
where $a$ is the axion field and $f_a$ is the axion constant.
 The axion field is compact too,  $a=0, \, \pm 2\pi, \, \pm 4\pi,..$  are identified. It is crucial that only one linear combination of $\varphi$ and $a$ acquires a mass, the orthogonal combination, 
\beq
A \equiv \varphi - x \,a\,,\qquad x= \frac{f_a^2}{\kappa^2}
\label{23}
\eeq
stays massless.\footnote{Exponentially small values of $x$ (i.e. exponentially small $f_a$)
must be excluded from our consideration since we need to maintain the axion-$\varphi$  mass much smaller than the masses
$m_{W}$ and $m_\beta$.
} Diagonalization transforms the Lagrangian (\ref{ququ}) into
\beqn
{\cal L}_{\rm eff}
&=& \frac{f_\Phi^2}{2}\,\left(\partial_\mu\Phi\right) \left(\partial_\mu\Phi\right) + 
 \frac{f_A^2}{2}\,\left(\partial_\mu A \right) \left(\partial_\mu A \right) +
\mu^3\, \cos  \Phi,\nonumber \\[3mm]
\Phi &=& \varphi + a\,,\qquad f_\Phi^2 =\kappa^2 \frac{x}{1+x}\,,\qquad  f_A^2 =f_a^2\, \frac{1}{x(1+x)}
\label{qu}\,.
\eeqn

A domain line can be built only out of the $\Phi$ field; the $A$ field cannot be excited inside the domain boundary strip 
because it is massless. There is no solution for $A$ other than $A$= constant (which can be put to zero) for all $y$ and $x$. 
The domain line solution is obtained from (\ref{14.35}) by the substitution
\beq
m_\varphi \to m_\Phi = \mu^{3/2} \,f_\Phi^{-1} = m_\varphi\sqrt{\frac{1+x}{x}}\,.
\eeq
Its tension (i.e., the tension of the $\Phi$ string) is
\beq
T_\Phi = 8 \mu^{3/2} f_\Phi = T_\varphi \sqrt{\frac{x}{1+x}}\,.
\eeq
Since across the domain line $\Delta \Phi = 2\pi $ and $A$ is not excited, and using 
\beq
\varphi = \frac{x\,\Phi + A}{1+x}\,,
\eeq
we conclude that 
\beq
\delta\varphi = 2\pi \frac{x}{1+x}\,.
\eeq

\begin{figure}[h]
\centerline{\includegraphics[width=12cm]{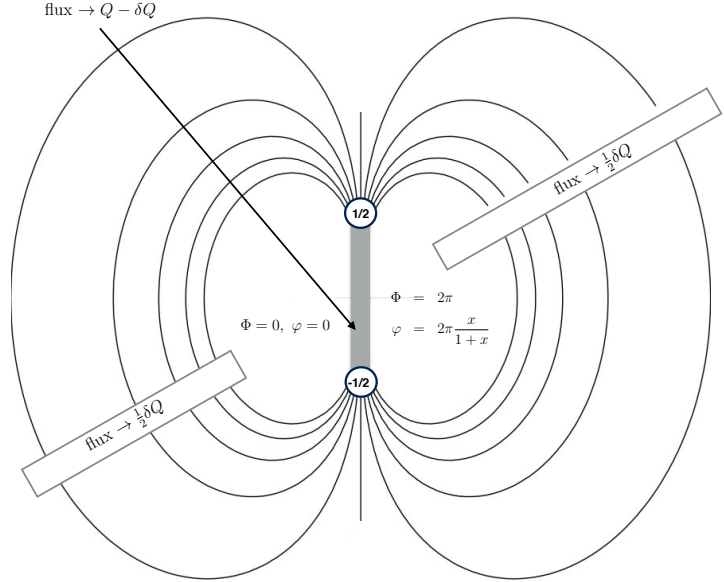}}
\caption{\small Electromagnetic dipole in there-dimensional compact electrodynamics. A part of the flux goes through the string (domain line), while the remaining flux is dispersed in a Coulomb-like manner.}
\label{tt}
\end{figure}
\noindent

Next, we observe that it is only the $\varphi$ field (or the photon $F_{0j}$, $j=1,2$) which interacts with the static
probe charge.
The minimal electric $U(1)$  charge $\frac 1 2$ is represented in the dual language by the $\varphi$ vortex with
$2\pi $ winding. Since in the presence of axion in the model the winding of the  $\varphi$ component of the vortex must be smaller than $2\pi$  the endpoint of the domain line will support the electric charge
\beq
Q- \delta Q \equiv \mbox{$ \frac 12$ - $\frac  12$} \frac{1}{1+x} \,.
\eeq
The remainder of the electric field flux, corresponding to $  \delta Q =
 \mbox{$\frac  12$} \frac{1}{1+x}$, is not squeezed inside the domain line (string),
but rather spreads out in a Coulomb-like manner (typical of a long electric dipole),
as shown in Fig. \ref{tt}. If $x \gg 1$ we return back to the Polyakov 3D string.
If $x\ll1$, the string dissolves. 

\section{Four-dimensional Yang-Mills on a cylinder} 
\label{4D YM}

In the previous sections, we considered the three dimensional Polyakov model.
In this section, we start from four dimensions and compactify the theory on $R^3\times S_L^1$, assuming that the circumference of the circle, $L$, is small, and then analyze the resulting theory in the low-energy limit. 
For definiteness the compactified direction will be aligned with the fourth axis $x^4$ which is taken to be a spatial direction.
The basic distinction from the previous case is the occurrence of two types of monopole-instantons: 
the first one is the same as in three dimensional theory,
while the second type of monopole-instanton is due to the nontrivial topology of $S^1_L$, namely $\pi_1(S^1) = Z$.

\subsection{Theory and perturbative analysis}

We consider $SU(2)$ Yang-Mills theory on $ R^3\times  S^1_L$   along with an axion  $a$:
\begin{eqnarray}\label{main action}
S=\int_{R^3 \times S_L^1}\, d^4 x
\left[
\frac{1}{4g^2}G_{mn}^aG_{mn}^a+\frac{F_a^2}{2}(\partial_m a)^2-i\frac{a}{32\pi^2}G_{mn}^a\tilde G_{mn}^a\right],
\end{eqnarray}
where $g$ and $F_a$ are the four-dimensional gauge and axion constants, $L$ is the $S^1$ circumference,
and the Latin letters run over $1,2,3,4$. Equation (\ref{main action}) refers to Euclidean space. Following \cite{Shifman:1979if}, we introduce the axion field 
$a$ via a heavy fermion $Q$ in the fundamental representation
coupled to a Higgs scalar $X$ singlet under $SU(2)$, 
\begin{eqnarray}\nonumber
{\cal L}_{Q+X }&=&i\bar Q D_n\gamma_n Q+\left(\bar Q_{L}Q_R X  +\mbox{H.c.}\right)\\[1mm]
&&+(\partial_n X)^2 -m_{X}^2 |X|^2+\frac{\lambda}{2}| X|^4\,.
\label{31}
\end{eqnarray}
The scalar field $X$ has two degrees of freedom, its modulus and phase, $$X=|X|\exp(i\alpha)\,.$$ With a judicious choice of parameters
the former will be very heavy and will determine the axion constant $F_a$ while the latter will be promoted to
the axion. 

The vacuum expectation of $X$ following from (\ref{31}) is\,\footnote{For simplicity we take $M_X$ to be real.}
\beq
|X| =\frac{m_X}{\sqrt{\lambda}}\,.
\eeq
By assuming  $m_X\gg1/L$ and $\lambda$ small, we ensure that the fermion is very massive and can be ignored
at energy scales much smaller than $1/L$, so it is irrelevant for what follows. However, the fermion loop produces the
coupling of the axion field to the gauge bosons which is not suppressed by the fermion mass, 
\beq
\Delta S = -\frac{i}{32\pi^2} \,a\,G_{mn} \tilde G_{mn}\,.
\label{33}
\eeq
 After dimensionally reducing the action (\ref{main action})  to $R^3$, we obtain 
\begin{eqnarray}
\nonumber
S_{\mbox{\scriptsize  3D}}&=&L\int_{R^3}\,d^3 x\left\{ \frac{1}{4g^2}G_{\mu\nu}^aG_{\mu\nu}^a+\frac{1}{2g^2}\left(\nabla_\mu \chi^a\right)\left(\nabla_\mu \chi^a\right)+\frac{F_a^2}{2}(\partial_\mu a)^2\right. \\[2mm]
&&\left. \quad-i\frac{a}{16\pi^2}\epsilon_{\mu\nu\rho}G_{\nu\rho}^a(\nabla_\mu \chi^a)
+{\cal V}[\Omega] 
\rule{0mm}{6mm}\right\},
\end{eqnarray}
where $\chi^a$ is the component of the gauge field along the compact (fourth) direction which in 3D acts as a compact adjoint scalar. 
In fact, upon compactifying the theory on $S^1_L$, one should sum up the tower of the Kaluza-Klein excitations of the gauge fields. This results in the Casimir potential 
given by the last term
\cite{Gross:1980br},
\begin{eqnarray}
{\cal V}(\Omega)=-\frac{2}{\pi^2 L^4}\sum_{n=1}^\infty\frac{|\mbox{Tr}\Omega^n|^2}{n^4}\,,
\end{eqnarray} 
where $$\Omega=\exp \left[{i\oint_{S_L^1}dx_4\, \chi}\right]$$ is the Polyakov line along $S^1_L$. 
Without loss of generality, we can perform a global $SU(2)$ transformation to align $\chi$ along the $\tau_3$ direction in the color space. Representing  $$\chi=\frac{v+\beta}{L}\,\frac{\tau_3}{2}\,,$$ where $v$ is the vacuum expectation value (VEV)  and $\beta$ is the field fluctuations, we find   
\beq
\langle\Omega\rangle=\mbox{diag}\left(e^{iv/2},\, e^{-iv/2}\right)\,.
\eeq
The potential ${\cal V}(\Omega)$ is minimized at $v=0$, and hence the center symmetry is maximally broken. 
$SU(2)$ gauge bosons are not Higgsed at $v=0$. 
This prevents the Abelianization due to $SU(2) \rightarrow U(1)$, which is essential for our  study of the theory using semi-classical methods. 

In order to force the Abelianization, we add a deformation $${\cal V}_{\mbox{\scriptsize def}}[\Omega]$$ to the theory \cite{Unsal:2008ch}. Such deformation can restore the center symmetry either fully or partially.  In the special case  when the potential is minimized at $\mbox{Tr}\Omega=0$ (or at $v=\pi$), the center symmetry is exactly preserved. This can be achieved by adding a double trace deformation
\begin{eqnarray}
{\cal V}^{\mbox{\scriptsize double trace}}_{\mbox{\scriptsize def}}= b |\mbox{Tr}\Omega|^2\,,
\end{eqnarray}
with some positive coefficient $b$. In this work we also consider the case  $$\mbox{Tr}\Omega \cong 0$$ (or $v \cong \pi$) which slightly shifts  us away from the exact center-symmetric vacuum.  As an example we will consider the following deformation:
\begin{eqnarray}
{\cal V}_{\mbox{\scriptsize def}}[\Omega]=\frac{\tilde b}{16L^4}\left|\mbox{Tr}\Omega\right|^4=\frac{\tilde b}{L^4}\cos^4\left(\frac{v}{2}\right)\,,
\end{eqnarray}
where in numerical calculation we set 
$$ \tilde b = 1000\,.$$
Then the total potential ${\cal V}+{\cal V}_{\mbox{\scriptsize def}}$ has two minima at $v\cong 3.105$ and $v\cong 3.178$. Thus, by adding a suitable deformation, the total potential is minimized at a non-zero expectation value of $v$, and  the $SU(2)\rightarrow U(1)$ breaking  takes place. In both situations (an exact or nearly-exact center symmetry) it is guaranteed that the $W$-bosons with mass $\frac{v}{L}$ are heavy,
provided we take the $S^1_L$ circle to be small, $L\Lambda_{\mbox{\scriptsize QCD}}\ll1$, where $\Lambda_{\mbox{\scriptsize QCD}}$ is the dynamical scale of the theory. This entails, in turn 
the freeze of the running of the coupling constant $g$ at a small value.  As a result, we are able to perform reliable semi-classical calculations. 

 Ignoring the heavy $W$-bosons, the resulting 3D Abelian action takes the form
\begin{eqnarray}
\nonumber
S_{\scriptsize U(1)\, {\rm 3D}}&=&L\int_{R^3}\, d^3 x\left\{ \frac{1}{4g^2}F_{\mu\nu}F_{\mu\nu}+\frac{1}{2g^2L^2}\left(\partial_\mu \beta\right)^2+\frac{F_a^2}{2}(\partial_\mu a)^2\right. \\[2mm]
&-& \left. i\frac{a}{16\pi^2 L}\epsilon_{\mu\nu\rho}F_{\nu\rho}\partial_\mu \beta
+{\cal V}[\Omega]+{\cal V}_{\mbox{\scriptsize def}}[\Omega]
\rule{0mm}{6mm}\right\},
\label{39p}
\end{eqnarray}
where $F_{\mu\nu}=G^{(3)}_{\mu\nu}$ and the superscript $(3)$ indicates the third direction in the color space.  

Next, we obtain a dual description of the three dimensional photon by introducing an
auxiliary term in the Lagrangian (cf. Sec. \ref{22}),
\begin{eqnarray}
\Delta S_\phi=\frac{i}{8\pi}\int d^3 x \epsilon_{\mu\nu\rho}\partial_\mu\phi F_{\nu\rho}\,.
\label{40}
\end{eqnarray}
Varying $\Delta S_\phi$ with respect to $\phi$, we obtain the Bianchi identity $\partial_\mu\epsilon_{\mu\nu\rho}F_{\nu\rho}=0$. Also,  varying $\Delta S_\phi+S_{\mbox{\scriptsize $U(1)$\,3D}}$ with respect to $F_{\mu\nu}$, we find
\begin{eqnarray}\label{duality}
F_{\nu \rho}=-\frac{ig^2}{4\pi L}\left(\partial_\mu \phi-\frac{a}{2\pi}\partial_\mu \beta  \right)\epsilon_{\mu\nu\rho}\,,
\end{eqnarray}
cf. Eq. (\ref{11}).
Substituting (\ref{duality}) in (\ref{39p}) and (\ref{40}),  we arrive at
\begin{eqnarray}
\nonumber
\Delta S_\phi+S_{\mbox{\scriptsize $U(1)$ 3D}}&\!=\!&\int d^3x \left\{ \frac{1}{2g_{\rm 3D}^2L^2}(\partial_\mu \beta)^2+\frac{f_a^2}{2}(\partial a)^2\right. \\[3mm]
&\!+\!&\left. \frac{g_{\rm 3D}^2}{32\pi^2 }\left(\partial_\mu\phi-\frac{a}{2\pi}\partial_\mu \beta\right)^2 +L{\cal V}[\Omega]+L{\cal V}_{\mbox{\scriptsize def}}[\Omega]\right\},
\end{eqnarray}
where we  defined the 3D coupling constant $$g_{\rm 3D}\equiv g/\sqrt{L}$$ and the 3D axion constant $$f_a\equiv \sqrt{L}F_a\,.$$
Since the potential $L{\cal V}[\Omega]+L{\cal V}_{\mbox{\scriptsize def}}[\Omega]$ is minimized at $v\cong \pi$,  the field $\beta$ acquires a mass $m_\beta\sim \frac{g}{L}$. Although $m_\beta$ is parametrically smaller than the $W$-boson mass, $\frac{v}{L}$, it is still exponentially larger than the photon mass (which is acquired non-perturbatively). Ignoring the massive fields, we find that the perturbative infrared Lagrangian is given by
\begin{eqnarray}
\label{pert infrared}
S_{\mbox{\scriptsize  pert }}=\int d^3x \left[\frac{\kappa^2}{2 }\left(\partial_\mu\phi\right)^2+\frac{f_a^2}{2}(\partial a)^2\right]\,,
\end{eqnarray}
and $\kappa=g_{\rm 3D}/4\pi$, cf. (\ref{ququ}).

\subsection{Non-perturbative contribution}
\label{npc}

In the previous section, we considered the infrared effective description of appropriately deformed Yang-Mills theory on $R^3\times S^1_L$ coupled to an axion. In this section, we take into account the instanton contribution
to generate a derivative-free axion coupling.  
  
Now, in addition to the 't-Hooft Polyakov monopole-instanton, we will have to deal with their Kaluza Klein excitations. 
  In fact, there are an infinite number of (anti)monopole-instantons  contributing to the partition function, thanks to the compact nature of  $S_L^1$. Fortunately enough, at weak coupling
  one has to take into account only the ones with the lowest action.  Using the dual photon description, the effective vertices of the  two main (anti)monopoles can be written as \cite{Unsal:2012zj}
\begin{eqnarray}
\nonumber
{\cal M}&=&\rho^3e^{-S_{{\cal M}}}e^{i\left(\phi+\frac{v a}{2\pi}\right)}\,,\quad \overline{{\cal M}}=\rho^3e^{-S_{{\cal M}}}e^{-i\left(\phi+\frac{v a}{2\pi}\right)}\\[3mm]
{\cal K}&=&\rho^3e^{-S_{{\cal K}}}e^{-i\left(\phi-\frac{v a}{2\pi}\right)}\,,\quad \overline{{\cal K}}=\rho^3e^{-S_{{\cal K}}}e^{i\left(\phi-\frac{v a}{2\pi}\right)}\,,
\end{eqnarray}
where the bar denotes anti-monopole. The actions of the monopoles are  
\beq
S_{{\cal M}}=\frac{4\pi}{g^2}v\,\,\,\mbox{ and} \,\,\, S_{{\cal K}}=\frac{4\pi}{g^2}(2\pi-v)\,,
\eeq
and  the pre-exponent $$\rho^3=\mbox{const}\times \frac{1}{L^3g^4}\,.$$  ${\cal M}$ is the conventional 't Hooft Polyakov monopole, while ${\cal K}$ is the twisted, or lowest Kaluza-Klein, monopole. The full effective action takes the form
\begin{eqnarray}\nonumber
S_{\mbox{\scriptsize eff}}&=&\int d^3 x\left\{  \frac{\kappa^2}{2 }\left(\partial_\mu\phi\right)^2+\frac{f_a^2}{2}(\partial_\mu a)^2\right. \\[2mm]
\label{reduced action}
&-&2\rho^3e^{-S_{{\cal M}}}
\left. 
\cos\left(\phi+\frac{va}{2\pi} \right)-2\rho^3e^{-S_{{\cal K}}}\cos\left(\phi-\frac{va}{2\pi}
 \right)\right\}.\label{full action a and phi}
\end{eqnarray}
 In the center-symmetric vacuum $v=\pi$, both ${\cal M}$ and ${\cal K}$ have the same action. Correspondingly,  the contribution from both monopoles have to be added with the same weight. 
 Then the mixing term on the $\phi$-$a$ mass matrix vanishes.

\subsection{\boldmath{$S_{{\cal M}}= S_{{\cal K}}$}}

If $S_{{\cal M}}=S_{{\cal K}} \equiv {\cal S}$, 
 then the potential in (\ref{reduced action}) takes the form
 \beq
 V= -4 \rho^3\,e^{-{\cal S}}\cos\phi\, \cos\frac{a}{2}\,.
 \label{pot47}
 \eeq
 It is obvious that the solution with pure $\phi$ domain wall  presented in Sec. \ref{sec23} and $a=0$ goes through. 
 If this solution is stable, then we can conclude that in this case axion's impact on confinement
 is absent.
 
\begin{figure}[h]
\centerline{\includegraphics[width=10cm]{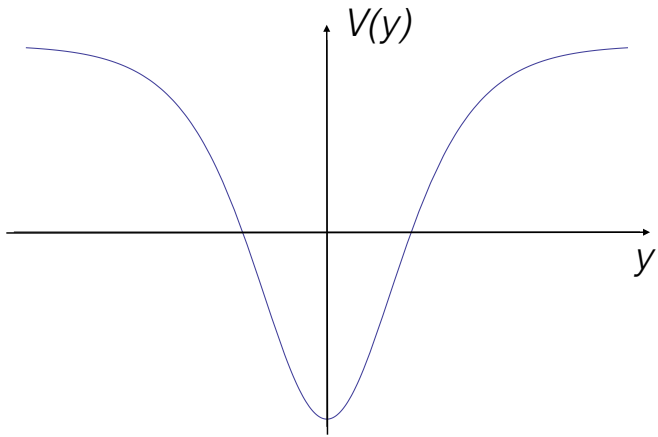}}
\caption{\small Potential in Eq. (\ref{48}).}
\label{opa}
\end{figure}

 The stability of the $a=0$ solution 
 can be checked by linearizing the equation for $a$ near $a=0$ and by determining the lowest energy eigenvalue.
 The equation is
 \beq
 \left(- d^2/dy^2 + \frac{\rho^3\, e^{-{\cal S}}}{f_a^2}\,\cos\phi_0(y)\right) a =\varepsilon_a a
 \label{48}
 \eeq
 (see Fig. \ref{opa} for the potential), 
 where $y$ is the direction perpendicular to the wall line and $\phi_0 (y) $ is the solution
 for the $\phi$ domain line discussed in Sec. \ref{sec23}. The lowest eigenvalue wavefunction must satisfy the boundary conditions $a (y=\pm\infty) =0$.

 To calculate the lowest eigenvalue, it is convenient to pass to dimensionless variables,
 \beqn
 \tilde y = m_\phi y\,, \qquad \tilde\varepsilon_a = \frac{\varepsilon_a}{m_\phi^2}\,, \qquad m_\phi^2
 =4\frac{\rho^3e^{-{\cal S}}}{\kappa^2}\,.
 \eeqn
 Then, (\ref{48}) becomes
 \beq
 \left(- d^2/d\tilde{y}^2 + \frac{1}{4x} \cos\phi_0(\tilde y)\right) a =\tilde\varepsilon_a a\,,
 \label{50}
 \eeq 
 where $x$ is defined in (\ref{23}).
 Numerical calculations yield that the lowest eigenvalue is positive for 
 $$x> \frac{1}{4}\,,$$ e.g.,  
 $$\tilde\varepsilon_{a\,\rm lowest} \cong 0.023$$ at $x=10$. The $a=0$ solution is stable at least
 locally. At $x=1/4$ one can solve Eq. (\ref{50}) analytically. One finds that at $x=1/4$ the lowest eigenvalue is exactly at $0$ and the zero eigenmode is
 \beq
 a_0=2\,\mbox{sech}\,\tilde y\,.
 \eeq
  For $x<\frac{1}{4}$, Eq. (\ref{50}) yields negative eigenvalues, e.g.,  $\tilde\varepsilon_{a\,\rm lowest} \cong -0.715$ at $x=0.1$, indicating the instability of the dual photon domain wall solution of Sec. \ref{sec23} with regards to generation of the $a$ field.

For values of $x<1/4$ we can start from the opposite side: we set $\phi=0$ and solve for $a$ to find the axion domain wall solution,
\beq
a_0 = 4\left[{\rm arcsin}\,{\rm tanh}\left(  m_a y
\right)+\frac{\pi}{2}
\right]\,, \quad m_a^2=\frac{\rho^3e^{-S}}{f_a^2}\,.
\label{axion DW}
\eeq
To check the stability of the solution (\ref{axion DW}), we linearize the equation of motion of $\phi$ near $\phi=0$ in the background of (\ref{axion DW}) to find the eigenvalue equation
\beq
 \left(- d^2/dy^2 + 4\frac{\rho^3\, e^{-{\cal S}}}{\kappa^2}\,\cos \left(\frac{a_0(y)}{2}\right)\right) \phi =\varepsilon_\phi \phi\,.
 \label{eigenvalue for phi}
 \eeq
Using the  dimensionless variables $$\tilde y=m_a y\,,\quad \tilde \varepsilon_\phi=\frac{\varepsilon_\phi}{m_a^2}$$ we obtain 
\beq
 \left(- d^2/d\tilde y^2 + 4x\,\cos \left(\frac{a_0(\tilde y)}{2}\right)\right) \phi =\tilde\varepsilon_\phi \phi\,.
 \label{dimless eigenvalue for phi}
 \eeq
Equation (\ref{dimless eigenvalue for phi}) is identical to Eq. (\ref{50}) upon the replacement $x\rightarrow 1/(16x)$. Thus, at $x<1/4$ the axion domain line makes  $\phi = 0$ locally stable because the lowest $\tilde\varepsilon_\phi >0$.

\begin{figure}[h]
\centerline{\includegraphics[width=9cm]{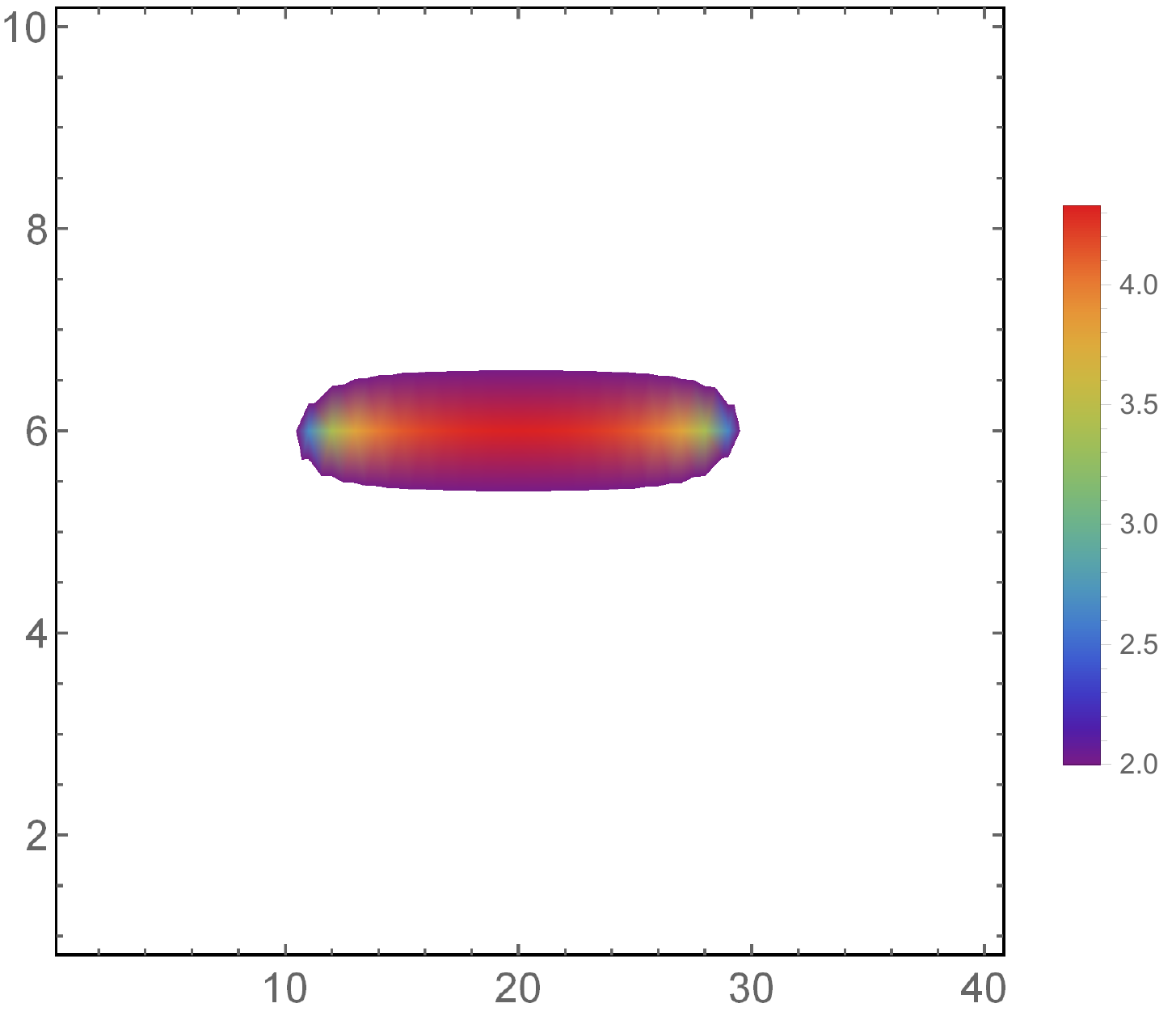}}
\caption{\small Numerical solution of the full equations of motion resulting from the action (\ref{full action a and phi}) in the case $S_{{\cal K}}= S_{{\cal M}}$.  Our simulations  in two dimensions are performed using the Gauss-Seidel relaxation method on a $40\times 40$ grid with periodic boundary conditions and two probe charges inserted at (10,6) and (30,6).  We plot the electric-field energy density $\frac{E^2}{2}=\frac{1}{2}\left[(\partial_1\phi)^2+[(\partial_2\phi)^2\right]$ for the parameters $\kappa=1$, $2\rho^3e^{-S_{{\cal K}}}=2\rho^3e^{-S_{{\cal M}}}=1$, and $x=0.005$. Even for such small values of $x$, we still can see the electric flux tube extending between the two probe charges.  Similar simulations for $x=0.005$ and $S_{{\cal K}}\neq S_{{\cal M}}$ show the dissolution of the electric flux tube.   }
\label{a phi DW}
\end{figure}

However, we know for sure that near the electric probe sources  $\phi \neq 0$.  This means that
 the lowest energy configuration has both components, a strongly modified  $\phi$ wall significantly
 different from that of
Eq. (\ref{sec23}), and a correspondingly modified $a$ wall.In this way confinement of charges will be maintained since the $2\pi$ periodicity in $\phi$ is preserved in the equations. This argument is confirmed by numerical analysis of the combined
$\phi$-$a$ wall in the potential (\ref{pot47}) at $x<1/4$, as shown in Fig. \ref{a phi DW}. The $\phi$ wall is clearly seen, with $2\pi$ vortices at the endpoints. 

\subsection{\boldmath{$S_{{\cal M}}\neq S_{{\cal K}}$}}
 
To recover the  situation in Sec. \ref{Axion's impact}, we
 need to destroy the equality of ${\cal M}$ and ${\cal K}$ contributions. 
There are two ways to suppress one of the monopoles and hence to make the situation parallel  with 
the case analyzed in Sec.  (\ref{Axion's impact}). First, we can shift the value of $v$ slightly from $v=\pi$.
In this way we  slightly depart from the exact center symmetry making $S_{{\cal K}}>S_{{\cal M}}$, or vise versa. Therefore, we can neglect either the ${\cal K}$ or ${\cal M}$ monopole contribution. This is a very good approximation in the small circle limit, $L\Lambda_{\mbox{\scriptsize QCD}}\ll1$, where we have $g^2\ll1$. 

The second option  preserves the center symmetry, but introduces a 4-D massless fermion in the fundamental representation of $SU(2)$. According to the index theorem on $R^3 \times S^1$ \cite{Poppitz:2008hr}, the fermion zero-mode will reside on one of the monopoles killing its contribution.

In both cases our action reduces to
\begin{eqnarray}
S_{\mbox{\scriptsize eff}}=\int d^3 x \left[\frac{\kappa^2}{2 }\left(\partial_\mu\phi\right)^2+\frac{f_a^2}{2}(\partial_\mu a)^2+\mu^3\cos\left(\phi\pm\frac{a}{2} \right)\right],
\end{eqnarray}
which is exactly the action analyzed in Sec. \ref{Axion's impact} after making a trivial shift $a/2\rightarrow a$.

\section{Axion in two-dimensional \boldmath{$CP(N-1)$} model}

From the pioneering Witten paper \cite{Witten78}, the large-$N$ solution of the two-dimen\-sional 
(nonsupersymmetric) $CP(N-1)$ model was found.
It was shown that the so-called $n$ fields ($N$-plets in the gauged formulation, see below) are confined, only 
$n\bar n$ mesons appear in the spectrum of the model. Later it was realized (e.g. \cite{Gorsky06})  that introducing the axion field one dramatically changes the spectrum of the model: confinement is eliminated: $n\bar n$ mesons
decay into their constituents (for a detailed discussion, see e.g., the review paper \cite{Gabadadze02}).
Here we will show that in fact, at large $N$, the above-mentioned mesons are very narrow, their decay rate
is suppressed by $\exp (-cN^\kappa)$ where $\kappa$ is a positive power, not necessarily integer. At $N=2$, however,  one can expect that the asymptotic  triplet states in the spectrum of the "axionless" model rapidly decay into the doublet states. 

First, we briefly review the model and then explain why the decay rate is exponentially suppressed. 

\subsection{\boldmath{$CP(N-1)$} in the gauged formulation}

 The Lagrangian of  $CP(N-1)$ model can be written as
\beq
{\cal L} = \frac{2}{g^2}\, \left[
\left(\partial_{\alpha} + i A_\alpha\right) n^*_{k}
\left(\partial_\alpha - i A_\alpha \right) n^{k}
-\lambda \left( n^*_{k} n^{k}-1\right)
\right]\,,
\label{onep}
\eeq
where $n^k$ is an $N$-component complex field,\footnote{Referred to as quarks or a soliton in
Ref.~\cite{Witten78}.} ($k = 1,2,...,N$) subject to the constraint
 \beq
n_{k}^*\, n^{k} =1\,.
\label{lambdaco}
\eeq
Moreovoer, $A_\mu$ is an auxiliary gauge field which has no kinetic term in the
bare Lagrangian.

The constraint (\ref{lambdaco}) can be implemented by the Lagrange multiplier $\lambda$
in (\ref{onep}). One could eliminate the field $A_\alpha$ in (\ref{lambdaco})
by virtue of the equations of motion,
\beq
A_\alpha =-\frac{i}{2}\, n^*_k \stackrel{\leftrightarrow}{\partial_\alpha} n^k\,.
\label{two}
\eeq
However, keeping in mind that a kinetic term for $A_\mu$  will be dynamically generated,
we will not use (\ref{two}).

Now, $g^2$ is a coupling constant; it is asymptotically free and defines
a dynamical scale of the theory $\Lambda$,
\beq
\Lambda^2 = M_{\rm uv}^2 \exp\left(-\frac{8\pi}{Ng^2_0}\right)\,,
\label{seven}
\eeq
where $M_{\rm uv}$ is an ultraviolet cut-off and $g^2_0$
is the bare coupling.

In the absence of axion, the solution of the   $CP(N-1)$ model
at large $N$  is determined by one loop and can be summarized as follows:
the constraint (\ref{lambdaco}) is dynamically eliminated so that all $N$ fields
$n^k$ become independent degrees of freedom with the mass  $\Lambda$.
The photon field $A_\mu$ acquires a kinetic  term
\beq
{\cal L}_{\gamma\,\,\rm kin}= -\frac{1}{4e^2} F_{\mu\nu}^2\,,\qquad e^2 = \frac{12\pi \Lambda^2}{N}\,,
\label{pki}
\eeq
and also becomes ``dynamical." We use quotation marks here
because in two dimen\-sions the kinetic term (\ref{pki})
does not propagate any physical degrees of freedom; its effect
reduces to an instantaneous
Coulomb interaction,
\beq
V_{\rm Coul} \sim \frac{\Lambda^2}{N}\,  |z|\,.
\label{35}
\eeq
Because of its linear growth
we get linear confinement acting between the $n,\,\,\bar n$ ``quarks."

\subsection{Axion's impact}

Now we switch on the axion,
\beq
{\cal L}_a = \frac 12 f_a^2\, (\partial_\mu a)^2 +\frac{a}{2\pi }\, \varepsilon_{\alpha\gamma}
\partial^\alpha A^\gamma\,,
\label{eight}
\eeq
where  $f_a$ is the axion constant.
In two dimensions it is dimensionless. We will start from the limit $f_a\gg1$, although this constraint is inessential.

Upon field rescaling, bringing kinetic terms to canonical normalization,
one obtains
\beq
- \frac{1}{4} F^2_{\mu\nu} + \frac{e}{2\pi f_a}\, a\, \varepsilon_{\alpha\gamma}
\partial^\alpha A^\gamma +
\frac 12 (\partial_\mu a)^2.
\label{dven}
\eeq
After diagonalization, the photon becomes massive and $a$ becomes its physical (propagating) degree of freedom.
The  mass is of order 
\beq
m_\gamma \sim f_a^{-1}\Lambda N^{-1/2}\,.
\eeq

Interaction between $n$ fields is now mediated by massive quanta, hence,
at  distances larger than $m_\gamma^{-1}$ the  confining potential (\ref{35}) is replaced by exponential fall off resulting in 
 deconfinement at distances $\gg m_\gamma^{-1}$.
 
 \begin{figure}[h]
\centerline{\includegraphics[width=9cm]{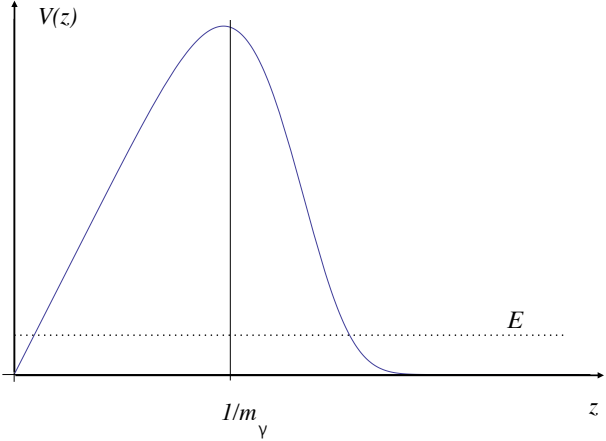}}
\caption{\small Coulomb interaction between $n$ and $\bar n$ cut off by the Yukawa exponent at distances
larger than $m_\gamma^{-1}$.}
\label{tt}
\end{figure}

The decay rate $w$ is determined by the Gamow mechanism. For nonrelativistic values of energy, $E\ll\Lambda$,
we obtain
\beq
w\sim \exp\left[-2\int \,dz\, \sqrt{\Lambda( V(z)-E)}\, 
\right]\sim \exp\left( -cf^{3/2}_aN^{1/4}\right),
\label{39}
\eeq
where $c$ is a numerical constant. As $f_a$ and $N$ decrease, the decay rate grows and becomes of order 1 at $f_a, N \sim 1$.

One can consider another mechanism of deconfinement.
In the ``axionless" $CP(N-1)$ model, there are
 $\sim N$ quasivacua  split in energy,
the splitting being
of order of $\Lambda^2/N$ (labeled by an integer $k$). Only the lower minimum is the true vacuum while
all others are metastable exited states.\footnote{In the large $N$ limit
the  decay rate is exponentially small, $\sim \exp(-N)$.}
 At large $N$,
the $k$ dependence of the energy density on the quasivacua, as well as the $\theta$ dependence, is
well-known
\beq
{ E}_k (\theta) \sim \, N\, \Lambda^2
\left\{1 + {\rm const}\,\left(\frac{2\pi k +\theta}{N}
\right)^2
\right\}
\,.
\label{split}
\eeq
At $\theta = 0$, the genuine vacuum corresponds to $k=0$,
while the first excitation corresponds to $k=-1$. At $\theta =\pi$,
these two vacua are degenerate, and at $\theta = 2\pi$
their roles interchange.

The energy split ensures
 kink confinement: kinks do not exist as  asymptotic states --- instead, they form
kink-antikink mesons. The regions to the left of the kink and to the right of the antikink
are the domains of the true vacuum (at $\theta =0$ it corresponds to $k=0$.)
The region between the kink and antikink
is an insertion of the adjacent quasivacuum  with $k=-1$.

When we introduce the axion, the vacuum angle $\theta$ is replaced by
a dynamical field, $a(t,z)$.
In the regions to the left of the kink and to the right of the antikink
$\langle a\rangle =0$.
If the region between the kink and antikink is large enough (this can happen e.g. if $m_a\sim \Lambda$),
the axion field in this region adjusts itself in such way  to minimize the energy,
$$
\langle a\rangle =0 \longrightarrow \langle a\rangle = 2\pi\,.
$$
The intermediate false vacuum decays in the true vacuum, through the axion wall formation
and restructuring of the $n$-field core in the middle. This probability can be estimated too,
\beq
\tilde w \sim \exp\left( -\tilde c N\right)
\eeq
and is smaller than (\ref{39}) at large $N$. 

\section{Conclusions}

We considered the impact of axions on confinement in two popular models.
In the three dimensional Polyakov model the mixing between the dual photon and axion is crucial
in changing the ``string" (domain line) structure. This change leads to a ``leakage" of a part of the electric flux to the
Coulomb regime. At small $f_a$ the domain line is built entirely from axions, and 
the electric flux disperses in the ``bulk."

In the two-dimensional $CP(N-1)$ model deconfinement disappears at $N=\infty$ and $f_a$ fixed.
However, if we fix $N$ and let $f_a$ become small, we observe the full blown deconfinement.

\section*{Acknowledgements}
We would like to thank Erich Poppitz and Mithat \"{U}nsal for comments. M.~S. is grateful to M.~Shaposhnikov and other colleagues from EPFL, where a part of this work was done.
The work of M.S.  is supported in part by DOE grant DE-SC0011842. The work of M.~M.~A. is supported by the Swiss National Science Foundation.

\vspace{5mm}

\end{document}